\documentclass[twocolumn]{aastex631}

\usepackage{hyperref}
\usepackage{soul}
\usepackage[T1]{fontenc}

\newcommand{\xmm}{{XMM-Newton}}

\newcommand{\swift}{{Swift}}

\newcommand{\nustar}{{NuSTAR}}

\newcommand{\ms}{\ensuremath{\rm M_{\odot}}}

\newcommand{\fluxcgs}{\ensuremath{\mathrm{erg}\,\mathrm{s}^{-1}\,\mathrm{cm}^{-2}}}
\newcommand{\lumcgs}{\ensuremath{\mathrm{erg}\,\mathrm{s}^{-1}}}

\begin{document}

\title{Constraint on the accretion of NGC 6946 X-1 using broadband X-ray data}

\correspondingauthor{Tanuman Ghosh}
\email{tanuman@rri.res.in}

\author[0000-0002-3033-5843]{Tanuman Ghosh}
\affiliation{Astronomy and Astrophysics,
Raman Research Institute,
Sadashivanagar, Bangalore 560080, India}

\author[0000-0003-1703-8796]{Vikram Rana}
\affiliation{Astronomy and Astrophysics,
Raman Research Institute,
Sadashivanagar, Bangalore 560080, India}



\begin{abstract}
We analyze broadband X-ray data of NGC 6946 X-1 and probe plausible accretion scenarios in this ULX. NGC 6946 X-1 is a persistent soft source with broadband continuum spectra described by two thermal disk components. The cool accretion disk temperature $\rm T_{cool} \sim 0.2$ keV and the presence of $\sim 0.9$ keV emission/absorption broad feature suggests the evidence of optically thick wind due to super-critical accretion. The hot geometrically modified accretion disk has an inner temperature of $\rm T_{hot} \sim 2$ keV with a radial dependent profile $\rm T(r) \propto r^{-0.5}$, expected in a slim disk scenario. Further, the measurement based on a realistic inclination angle of the disk indicates that the mass of the host compact object is comparable to $\rm \sim 6-10 ~\ms$ non-rotating black hole or the system hosts a moderately magnetized neutron star with $\rm B \lesssim 2 \times 10^{11}$~G magnetic field. Overall, the detected spectral curvature, high luminosity, flux contribution from two thermal disk components, and estimated accretion rate imprint the super-Eddington accretion scenario.

\end{abstract}



\section{Introduction} \label{sec:intro}
Ultraluminous X-ray sources (ULXs; $\rm L_x>10^{39}$ \lumcgs) are best understood to be the super-Eddington accreting X-ray binaries predominantly abundant outside of our Galaxy. Discovery of the extragalactic neutron star ULXs \citep{Bachetti2014, Furst2016N, Israel2017F, Israel2017M, Brightman2018, Carpano2018, Rodriguez2020, Sathyaprakash2019} firmly established that the Eddington ratio ($\rm \lambda_{Edd} = L_x/L_{Edd}$) can reach a factor of few hundreds. Broadband spectral analyses have shown that two-component thermal disk emission and an additional component like coronal comptonization or emission from a magnetized accretion column provide the X-ray continuum (see, e.g., \citealt{Walton2018apr, Walton2020}) in most of the sources.

The thermal photons work as the seed photons for the Compton up-scattering process, which give harder photons $\sim 20$ keV in many ULX sources. These different emission components are highly dependent on the geometry of the disk, which determines the hardness of the spectra. The softest of ULXs are assumed to be viewed close to the plane of the disk. Hence, the hot inner regions are obscured out of the line of sight. Due to the high mass accretion rate in the case of super-Eddington accretion, the inner region of the disk geometrically deviates from the standard thin disk. It becomes a slim accretion disk where the disk scale height is comparable to the radius, i.e., $\rm h \approx r$ \citep{Abramowicz, Sadowski, Dotan}. Simplistically, such a geometrically modified disk is often ascribed to the powerlaw temperature profile of $\rm T(r) \propto r^{-p}$, where $\rm p \sim 0.5$ for a geometrically slim disk \citep{Sadowski} and at the limit of $\rm p = 0.75$, it takes the form of standard thin \citet{Shakura1973} disk. Due to super-Eddington accretion, near the spherization radius, wind is generated, and it outflows the material from the inner region of the disk.

The two thermal disk components in ULXs have several interpretations. For example, in a black hole system, the soft/cool disk component is ascribed to the reprocessed photons from optically thick wind, and the hard counterpart comes from the inner accretion flow (e.g., \citealt{Walton2014, Walton2015, Luangtip2016}). Weakly magnetized ($\rm B<10^{11}$ G) neutron stars have a similar accretion scenario to that of the super-Eddington black holes \citep{King2016}. One possible explanation is that the thin \citep{Shakura1973} disk imprints the cool thermal component, and the optically thick plasma at the boundary layer of the neutron star emits harder spectra \citep{Syunyaev1986, Koliopanos2017}. Even for highly magnetized neutron stars, due to high luminosity ($\rm L_x \gtrsim 5 \times 10^{39}$ \lumcgs), the material within Alfv\'en radius becomes optically thick and emits hard photons, whereas the cool component comes from the truncated accretion disk \citep{Koliopanos2017, Mushtukov2017}. An alternative scenario is that the soft emission component in ULXs, which resembles the cool accretion disk-like emission, comes from the optically thick wind expected for super-critical accretion.

The study by \citet{Qiu2021} on a sample of ULX sources showed that the soft thermal component is a signature of optically thick wind because of the constant blackbody luminosity. In that study, it was established that this correlation between wind scenario and cool blackbody-like emission is plausible in both black holes and pulsar ULXs. The higher blackbody luminosity than the Eddington limit of neutron stars could be explained by the reduced scattering cross-section in the presence of a high magnetic field or increased radiation due to magnetic buoyancy \citep{Qiu2021}.

The wind or outflow also imprints the emission and absorption lines in the ULX spectra, especially around $\sim 1$ keV. These strong line features have been seen in high-quality grating spectral data in several ULXs, e.g.,  NGC 55 ULX, NGC 247 ULX-1, NGC 1313 X-1, NGC 5408 X-1, NGC 300 X-1, Ho IX X-1, Ho II X-1 (see, e.g., \citealt{Pinto2016, Pinto2017, Pinto2020, Pinto2021, Kosec2018feb, Kosec2018sep, Kosec2021}). Interestingly, it has been found that the soft ULX sources are better candidates for detecting these outflow features compared to the hard sources and can be related to the strong wind/outflow and favorable geometrical occultation of the inner flow of the disk.

NGC 6946 galaxy (distance $\sim 7.72$ Mpc; \citealt{Anand2018}) is the host of multiple ULXs \citep{Earnshaw2019}. NGC 6946 X-1 (RA:20:35:00.7, DEC:+60:11:31) is a soft ULX source often referred to as ULX-3 (see \citealt{Earnshaw2019} and references therein), has shown emission line signatures \citep{Pinto2016, Kosec2021}, a similar feature shown by several other soft ULX sources. We study its broadband spectral properties using \xmm\ and \nustar\ data and provide some crucial constraints on the accretion mechanism in this ULX from the continuum features.

In \S~\ref{sec:Data}, we discuss the data utilized in this paper and their extraction processes. \S~\ref{sec:Results} describes the analysis of the data and the obtained results. Finally, we discuss and conclude our findings in \S~\ref{sec:Discussion}.

\begin{deluxetable*}{cccccc}
\tablenum{1}
\tablecaption{Observation log of NGC 6946 X-1 \label{tab:logtable}}
\tablewidth{0pt}
\tablehead{
\colhead{Serial No.} & \colhead{Observation ID} & \colhead{Observation ID} & \colhead{Observation start date} & \colhead{Epoch ID} &
 \colhead{Spectral Exposure time (ks)}  \\
\nocolhead{} & \colhead{\xmm} &  \colhead{\nustar} &
\nocolhead{} & \nocolhead{} & \colhead{pn/MOS1/MOS2/FPMA/FPMB}
}
\startdata
1 & 0870830101 & 50601001002 &  2020-07-08 & XN1 & 12.5/16/16/100/99  \\
2 & 0870830201 & 50601001004 &  2020-12-13 & XN2 & -/16/16/94/91  \\
3 & 0870830301 & 50601001006 &  2021-04-02 & XN3 & 8.5/14/14/83/84 \\
4 & 0870830401 & 50601001008 &  2021-05-25 & XN4 & 12/-/16/88/89 \\
\enddata
\tablecomments{pn data are not used for observation 0870830201 and MOS1 data are not available for observation 0870830401.}
\end{deluxetable*}

\section{Data} \label{sec:Data}
We analyze the simultaneous broadband data of NGC 6946 X-1 jointly observed by \xmm\ and \nustar\ in 2020-2021. The source has been previously analyzed on several occasions by archival \xmm\ and \nustar\ data (see, e.g., \citealt{Kajava2009, Hernandez, Middleton2015, Pintore2017, Kosec2018feb, Earnshaw2019, Kosec2021, Qiu2021}) and the readers are directed to these papers for an overview of the previous analyses. Here, we focus on the recent joint observations by \xmm\ and \nustar\ observed in 2020-2021. Previous joint \xmm\ and \nustar\ observation data taken in 2017 were analyzed in \citealt{Earnshaw2019} in detail. Therefore, we compare the results from 2020-2021 broadband data to those obtained in \citealt{Earnshaw2019}. The observations used in this paper are tabulated in Table \ref{tab:logtable}.

The \xmm\ data are processed by SASv20.0.0, and the EPIC products are extracted using \texttt{epproc} and \texttt{emproc} tools. In observation 0870830201, the source falls in the chip gap of the pn camera, thus significantly affecting the data. Therefore, for this observation, we do not utilize the pn data for the scientific analysis. In observation 0870830401, MOS1 data are not available. The data are cleaned from background flaring by \texttt{espfilt} task, and \texttt{evselect} tool is used to generate the spectra and light curves from the cleaned event files with \texttt{PATTERN<=4} for pn and \texttt{PATTERN<=12} for MOS data. \texttt{rmfgen} creates the redistribution matrix files and \texttt{arfgen} is used to create the ancillary response files for the source spectrum. The source region is selected from a $20$ arcsec circular region, and the background is from a $40$ arcsec circle in a nearby region from the same chip. We select \texttt{FLAG==0} criteria for extraction of spectra for all cameras. The spectra are grouped to have $20$ counts per bin with an oversampling factor of $3$.

The \nustar\ data are extracted using HEASOFT version 6.31. The raw data are cleaned and pre-processed using \texttt{nupipeline} with \texttt{saacalc=3}, \texttt{saamode=STRICT} and \texttt{tentacle=yes} parameters to maintain a conservative approach while handling the background due to South Atlantic Anomaly. The \texttt{nuproducts} task is further utilized to generate the spectra and lightcurves from the cleaned data. The source extraction region is a $40$ arcsec radius circle, and a nearby $60$ arcsec radius circle from the same chip is chosen for the background region for all observations. The spectra are grouped to have $20$ counts per energy bin.

\section{Analysis and Results} \label{sec:Results}

We first individually analyze the joint \xmm\ and \nustar\ spectra for all four epochs using XSPEC v12.13.0b \citep{XSPEC}. Throughout the paper, we have estimated uncertainties in parameters within a 90\% confidence interval unless mentioned otherwise. We have used updated abundance \citep{Wilms} and cross-section \citep{Verner} for the neutral absorption model \texttt{tbabs}. The flux is measured using the convolution model \texttt{cflux}. The \xmm\ spectra are utilized in the $0.3-10.0$ keV energy range, whereas the \nustar\ spectra are fitted in the $3.0-20.0$ keV energy range, beyond which background significantly dominates the spectra.

\begin{figure}[t]
\centering
    \includegraphics[width=1.\linewidth]{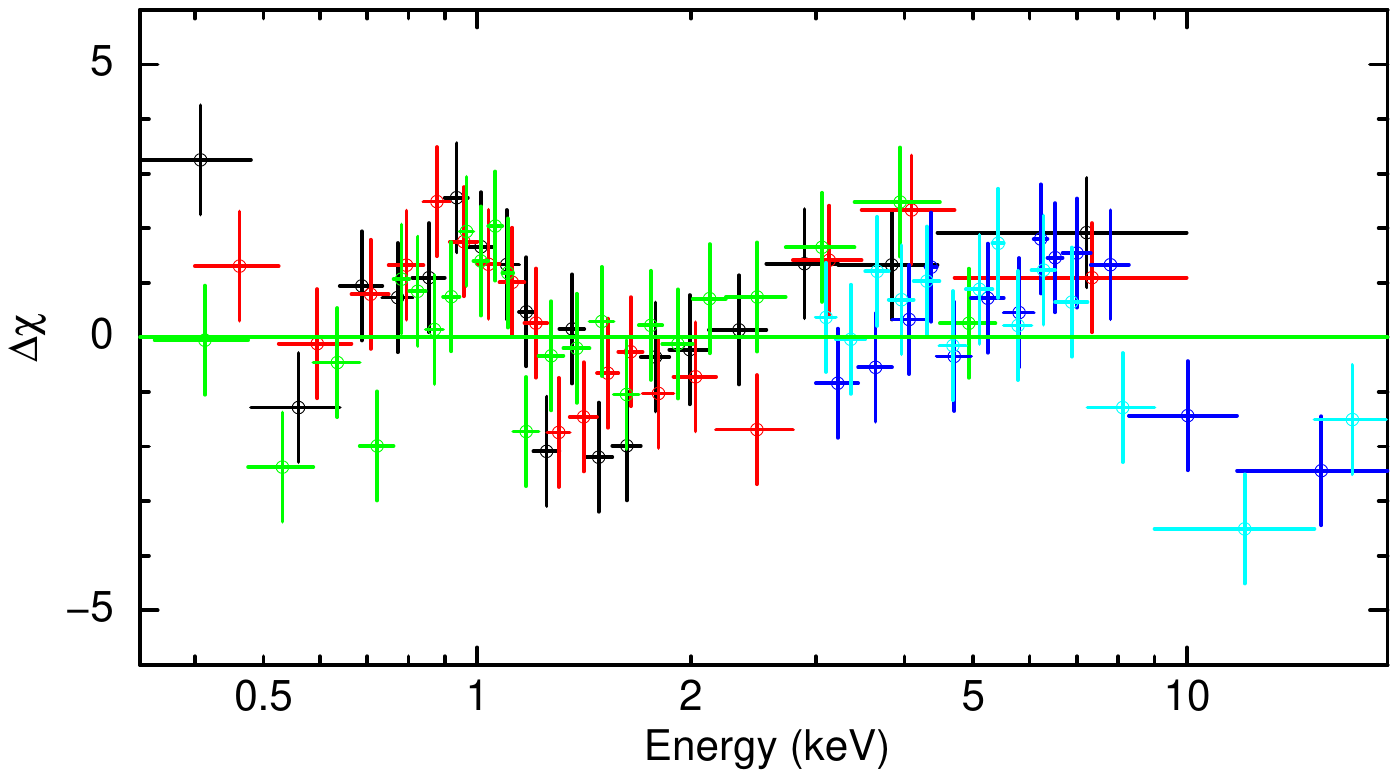}
\caption{The residual from absorbed powerlaw fit of XN1 broadband data shows the presence of $\sim 0.9$ keV broad hump feature and a high energy spectral turnover. Data have been rebinned for visual purposes.
\label{fig:powerlaw_residual_XN1}}
\end{figure}

\begin{figure}
\centering
    \includegraphics[width=1. \linewidth]{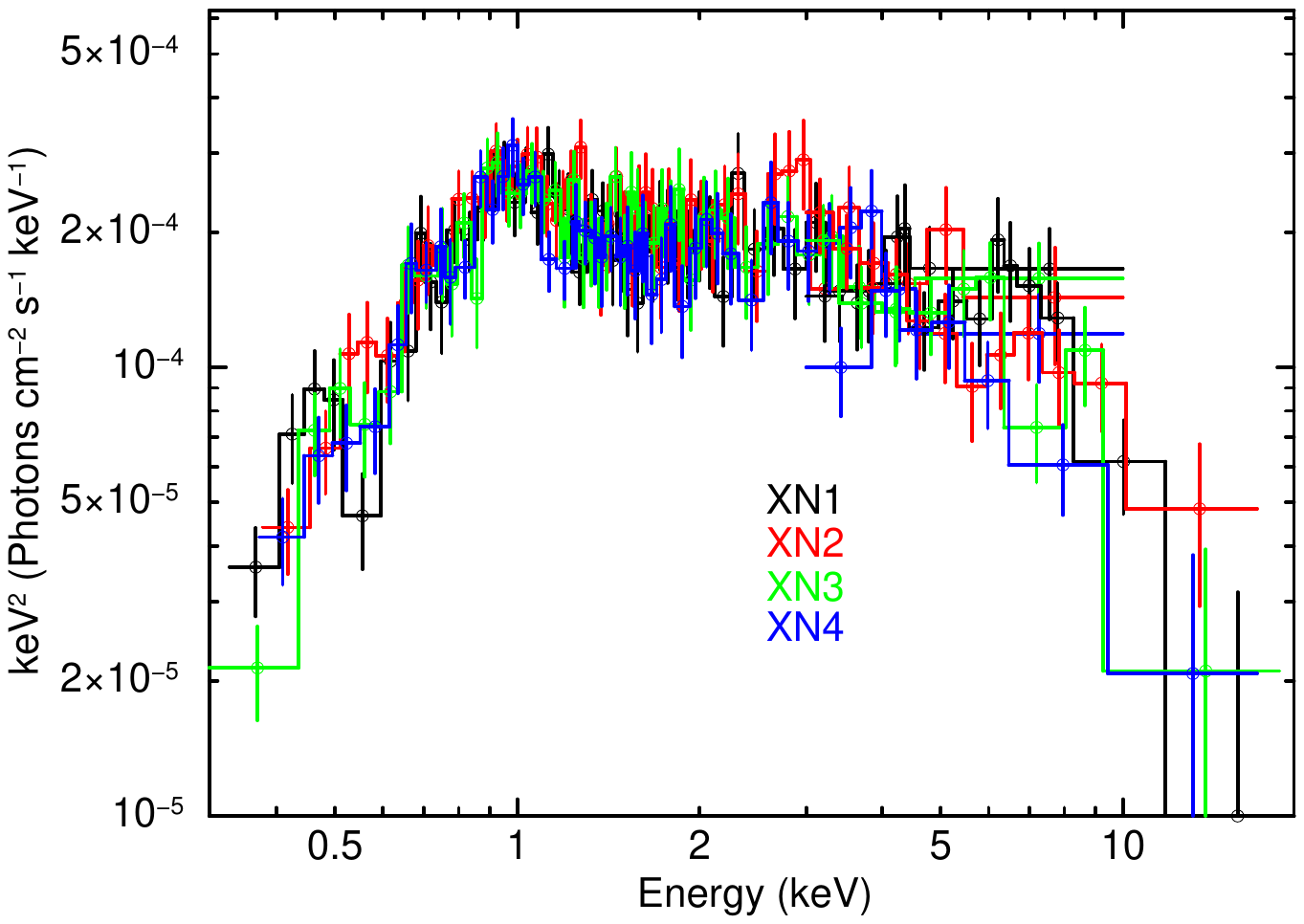}
\caption{Unfolded spectra using \texttt{powerlaw} model of 0 index and arbitrary normalization. For clarity, only MOS2 and FPMA spectra are shown for all four epochs. Visually, it is apparent that all four epochs exhibit overlapping spectral features. Data have been rebinned for visual purposes.
\label{fig:spectra_eeufspec}}
\end{figure}

First, we implement an absorbed \texttt{powerlaw} model and find that in all observations, there is a broad feature in the residual around $\sim 0.9$ keV. However, the measurement uncertainties on the parameters of the broad feature varies depending on the signal-to-noise ratio (S/N) of the data. We include a \texttt{gaussian} model component to justify this soft broad feature following the method by \citealt{Earnshaw2019}. The cross-calibration constant for MOS2 is fixed to 1 and left free to vary for other detectors. The broadband data indicate the typical powerlaw break in the spectra like other ULXs. In figure \ref{fig:powerlaw_residual_XN1}, we show the residual for an absorbed \texttt{powerlaw} fit for the XN1 epoch. The residual indicates an apparent broad feature around $\sim 0.9$ keV and a high energy spectral cutoff. Hence, we fit the spectra for all epochs with an additional multiplicative component \texttt{highecut} on top of the \texttt{powerlaw} continuum. Therefore, an absorbed \texttt{gaussian+highecut*powerlaw} model gives a good fit in all cases. We find congruent parameter ranges throughout all observations of 2020-2021, implying that the source does not significantly vary in spectral nature during these observations. From the unfolded spectra, it is visually apparent that broadband spectra from these four epochs possess similar features (see figure \ref{fig:spectra_eeufspec}). However, it is important to note that some parameters are not properly constrained for the XN2 epoch due to the unavailability of pn data but remain consistent within the 90\% confidence range of the parameters from other epochs. Hence, we simultaneously fit the spectra for all 2020-2021 epochs with the same model with parameters for all epochs linked.

While simultaneously fitting all these spectra with an absorbed \texttt{powerlaw}, we get a $\rm \chi^2/d.o.f \simeq 858/653$. When we add a \texttt{gaussian} component, the improved fit statistics is $\rm \chi^2/d.o.f \simeq 705/650$. Nevertheless, an additional \texttt{highecut} component improved the statistics further. However, we find that with the addition of \texttt{highecut} component, the neutral absorption ($\rm N_H$) gives a best-fit value close to the Galactic absorption value  $\rm N_H \sim 0.22 \times 10^{22}$ cm$^{-2}$ \citep{HI4PI}. This is apparently coming from the fact that NGC 6946 is a face-on galaxy and the local absorption is less compared to the sensitivity of the data; hence, we fix the $\rm N_H$ to the Galactic value throughout the analysis. We find that the cutoff and folding energies are around $\rm E_{cut} \sim 6.34^{+0.72}_{-0.63}$ keV and $\rm E_{fold} \sim 4.06^{+1.23}_{-1.06}$ keV with a powerlaw index of $\rm \Gamma \sim 2.35^{+0.03}_{-0.04}$, portraying it as a soft source (see table \ref{tab:table_highcut_pow}). The $\rm \chi^2/d.o.f$ is $556/649$ for this simultaneous fit. Finally, we study this source by simultaneously fitting all these four epochs spectra for different spectral models by linking the spectral parameters. This helps constrain the individual parameters with much better precision and, in turn, helps constrain the physical parameters we estimate from the analysis.

In this paper, we mainly focus on the continuum spectral fitting of 2020-2021 broadband X-ray observations and undertake a simpler approach to fit the $\sim 0.9$ keV feature by a Gaussian model (see discussion on such a similar feature in other ULXs, e.g.,\citealt{Middleton2014, Middleton2015, Ghosh2022NGC4395}). We investigate the continuum fitting with different models relevant to the physical scenarios expected in ULX systems. A single component \texttt{diskpbb} model does not provide a good fit (\rm $\chi^2/d.o.f \simeq 1340/654$). There are significant residuals in the soft energy range where typically soft disk blackbody component and the $\sim 0.9$ keV broad hump-like feature dominate the spectra. Hence, we utilize the well-explored two-component thermal disk models like \texttt{diskbb+diskpbb} to fit the continuum. A Gaussian and this two-component thermal disk continuum provide a statistically acceptable fit (see Table \ref{tab:table1}).

\begin{figure}
\centering
    \includegraphics[width=1. \linewidth]{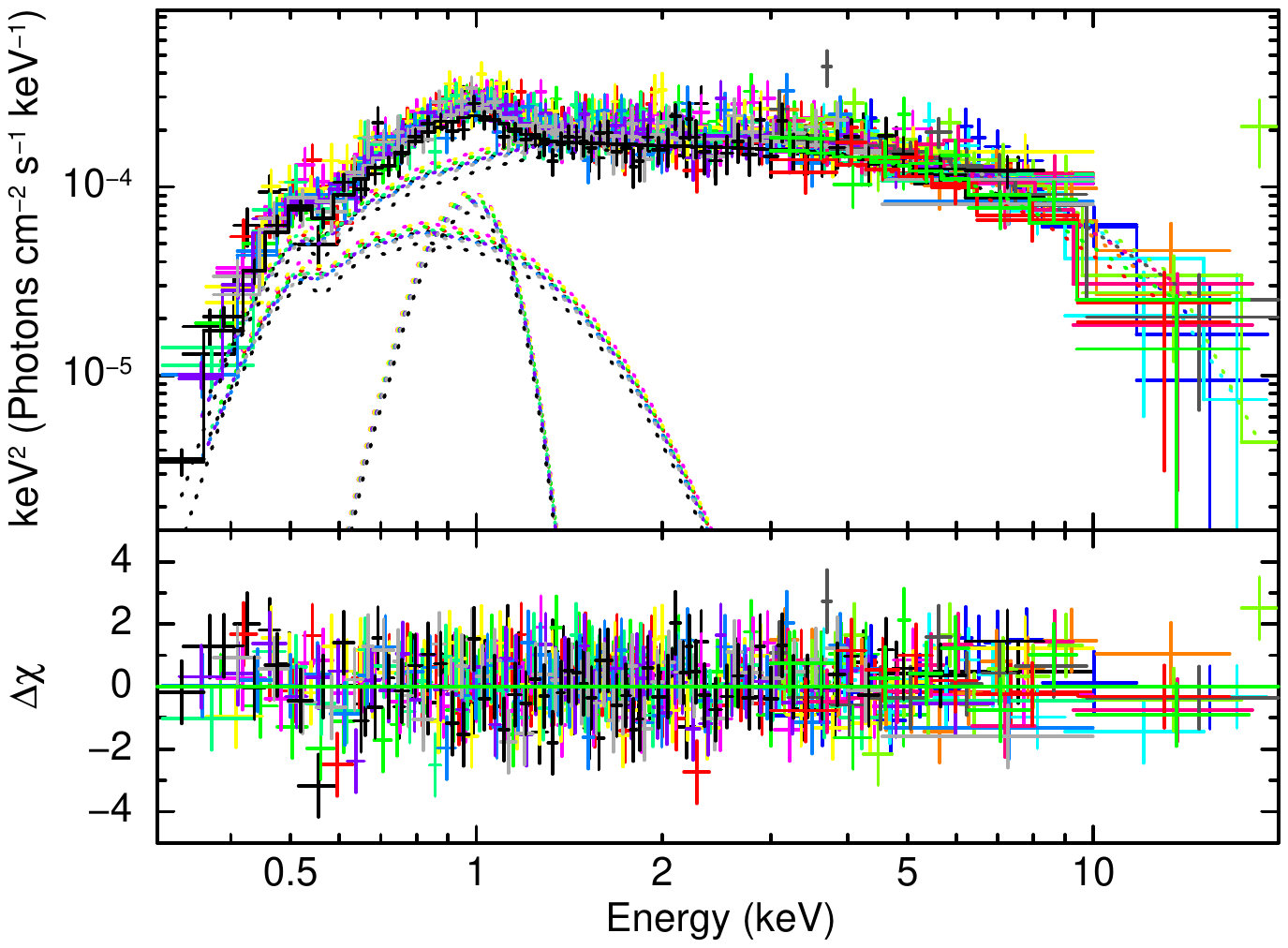}
\caption{The spectra, model components, and residuals are shown for all epochs simultaneous fit with \texttt{{tbabs*(gaussian+diskbb+diskpbb)}} model. Data have been rebinned for visual purposes.
\label{fig:residuals_spectra}}
\end{figure}

When we fit the \texttt{tbabs*(gaussian+diskbb+diskpbb)} model, we get the cool disk temperature around $\sim 0.22$ keV and the hot disk temperature around $\sim 2$ keV. The radial dependence of temperature parameter `p' in \texttt{diskpbb} component converges to the hard limit of $0.5$, i.e., the slim-disk limit. Hence, we fix this parameter to $0.5$. We find that no additional hard component is required for the $0.3-20.0$ keV spectra in any epoch. The model parameters and flux are described in Table \ref{tab:table1}. The spectra, models, and residuals are shown in figure \ref{fig:residuals_spectra}.

We analyse light curves of NGC 6946 X-1 to search for any pulsation in the source. We implement acceleration search technique by utilizing the HENDRICS \citep{HENDRICS, Stingray} tool \texttt{HENaccelsearch} to search for pulsation in the frequency range of $0.01-6.8$ Hz in \xmm\ data and $0.01-10.0$ Hz in \nustar\ data. However, there is no significant detection of pulsation in any of the observations.

\begin{deluxetable}{CCC}
\tablenum{2}
\tablecaption{ Parameters for the fitted \texttt{tbabs*(gaussian+highecut*powerlaw)} model of NGC 6946 X-1 broadband spectra. The total absorbed flux ($\rm F_x$) is measured in the $0.3-20.0$ keV energy range. \label{tab:table_highcut_pow}}
\tablewidth{0pt}
\tablehead{
\colhead{Parameter} & \colhead{Unit} & \colhead{Parameter values} \\
}
\startdata
\hline
\rm   N_H & 10^{22} \rm cm^{-2} & 0.22 (\rm fixed)  \\
\rm   E_{line} & \rm keV & 0.89 \pm 0.03   \\
\rm   \sigma_{line}  & \rm keV & 0.15^{+0.03}_{-0.02}  \\
\rm   Norm & 10^{-5} photons ~cm^{-2} ~s^{-1} & 6.62^{+1.55}_{-1.20}  \\
\rm   E_{cut} & \rm keV & 6.34^{+0.72}_{-0.63}  \\
\rm   E_{fold} & \rm keV  & 4.06^{+1.23}_{-1.06}  \\
\rm   \Gamma &  & 2.35^{+0.03}_{-0.04}  \\
\rm   Norm_{pow} & 10^{-4} & 2.61 \pm 0.14 \\
\rm  \chi^2/d.o.f &  & 556/649 \\
\rm   F_{x} & 10^{-13} ~\rm erg ~\rm cm^{-2} ~\rm s^{-1} & 8.24 \pm 0.41 \\
\enddata

\end{deluxetable}

\begin{deluxetable}{CCC}
\tablenum{3}
\tablecaption{Parameters for the fitted \texttt{tbabs*(gaussian+diskbb+diskpbb)} model of NGC 6946 X-1 broadband spectra. The bolometric unabsorbed total flux and the flux from individual disk components are measured in the $0.01-100.0$ keV energy range. \label{tab:table1}}
\tablewidth{0pt}
\tablehead{
\colhead{Parameter} & \colhead{Unit} & \colhead{Parameter values} \\
}
\startdata
\hline
\rm N_H & 10^{22} \rm cm^{-2} & 0.22 (\rm fixed)  \\
\rm E_{line} & \rm keV & 0.92 \pm 0.03   \\
\rm \sigma_{line}  & \rm keV & 0.13 \pm 0.03  \\
\rm Norm &  10^{-5} photons ~cm^{-2} ~s^{-1} & 4.84^{+1.78}_{-1.20}  \\
\rm T_{thin} & \rm keV & 0.22^{+0.02}_{-0.03}  \\
\rm Norm_{thin} &  & 8.48^{+7.22}_{-2.97}  \\
\rm T_{slim} & \rm keV & 2.03^{+0.13}_{-0.11}  \\
\rm p &  & 0.5 (\rm  fixed)  \\
\rm Norm_{slim} & 10^{-4} & 4.45^{+1.31}_{-1.05} \\
\rm \chi^2/d.o.f &  & 555/649 \\
\rm  F^{bol}_{unabs} &  10^{-12} ~\rm erg ~\rm cm^{-2} ~\rm s^{-1} & 2.61 \pm 0.14 \\
\rm F^{bol}_{diskbb} & 10^{-13} ~\rm erg ~\rm cm^{-2} ~\rm s^{-1} & 4.17^{+0.61}_{-0.64} \\
\rm F^{bol}_{diskpbb} & 10^{-12} ~\rm erg ~\rm cm^{-2} ~\rm s^{-1} & 2.12 \pm 0.12 \\
\enddata

\end{deluxetable}

\section{Discussions and Conclusions}\label{sec:Discussion}
We discuss the broadband spectral properties of a soft ULX source NGC 6946 X-1 from the 2020-2021 observations. Previous studies (e.g., \citealt{Pinto2016, Kosec2021}) have detected emission lines from the high-resolution grating spectra and confirmed the presence of wind/outflow in the system. \citealt{Earnshaw2019} studied its first broadband spectral properties using \xmm, \nustar, and \swift\ data. They found that X-1 is a persistent ULX as its flux remained consistent with previous estimates (e.g., \citealt{Middleton2015}). Interestingly, our analysis finds that even in 2020-2021 observations, X-1 shows flux $\rm F_x \simeq (8.0 \pm 0.4) \times 10^{-13}$ \fluxcgs~ in $0.3-10.0$ keV energy range, which is close to the previous findings. Hence, we discern that this source is indeed a persistent and steady ULX. One crucial comparison with the analysis of \citealt{Earnshaw2019} for the 2017 \xmm +\nustar\ observation is that measurement of $\rm N_H$ is slightly higher in that observation. Even though our choice of Galactic absorption is higher than that used in \citealt{Earnshaw2019}, we find that the data can constrain an additional local absorption on top of the Galactic absorption in that dataset. Another important result is that the simultaneous fitting of these recent 2020-2021 data properly constrains the presence of spectral curvature in this ULX, which is an important feature of super-Eddington accretion in ULXs. We further discuss the implications of the spectral fittings and their relevance to the physical scenario in the ULX. Notably, these models are phenomenological, and any degeneracy between different model combinations can have a considerable impact on the observational implications in the source spectra. With this caution, we primarily discuss the broadband spectral features and results from the thermal disk continuum models. The $\sim 0.9$ keV broad emission line feature, studied for several ULXs, has been simply treated here with a Gaussian model. We focus on the continuum properties and discuss some physical aspects related to that.

\subsection{Outflow scenario in super-critical accretion}
There are ULX sources, e.g.,  NGC 247 ULX1, NGC 55 ULX, NGC 4395 ULX1, NGC 1313 X-1, NGC 5408 ULX1, NGC 300 X-1, Ho IX X-1, Ho II X-1, and NGC 5408 X-1 \citep{Middleton2015dec, Pinto2016, Pinto2017, Earnshaw2017, Pinto2021, Ghosh2022NGC4395} where the strong $\sim 0.9$ keV feature is detected. In fact, most of the bright ULXs where strong $\sim 0.9$ keV line emission/absorption features have been discovered are found to have predominantly softer spectra \citep{Kosec2021}. A similar $\sim 0.9$ keV feature is seen in NGC 6946 X-1 (see also \citealt{Middleton2015dec, Pinto2016, Kosec2021}). This feature is assumed to be related to the wind or outflow in super-Eddington accretion scenario \citep{Takeuchi}. Mostly soft sources show these lines because the line of sight inclination is close to the disk plane, and we observe the inner hot photons only after being down-scattered by the wind, which eventually appear to us as soft photons. Also, due to such inclination, the inner disk is occulted by the wind clouds, and we receive a higher fraction of line emission/absorption from the winds. The strength and energy of the lines often depend on the wind velocity and the direction of its motion. In NGC 6946 X-1, we see a moderate fraction of both soft and hard photons. Generally, the hard sources are understood as close to the face-on system, whereas the ultra-soft sources are understood to be viewed close to edge-on. Hence, the accretion disk in NGC 6946 X-1 can be interpreted as moderately inclined towards the line of sight (see also \citealt{Pinto2017}). As discussed earlier, it is widely accepted that the cool disk component in ULXs comes from the optically thick wind due to super-Eddington accretion. We obtain that the bolometric luminosity of the cool disk component is $\rm L_{bol} = 4\pi D^2 F_{bol} \simeq 3 \times 10^{39}$ \lumcgs, which is consistent with a marginally above Eddington luminosity of a $\sim 10$ \ms black hole and a super-Eddington luminosity for a neutron star system. Hence, we can interpret that the soft spectral components like the cool accretion disk and the broad $\sim 0.9$ keV feature are related to the emission from optically thick wind due to accretion close to or above the Eddington accretion rate. The hard spectral component, described by a hot \texttt{diskpbb} model, can originate from the inner accretion flow for a black hole or neutron star ULX system.

The \texttt{diskpbb} model with the temperature profile of $T(r) \propto r^{-0.5}$ signifies significant advective cooling and photon trapping in a slim accretion disk. In a black hole accretion scenario, it is understood that the inner radius of the hot slim disk is the inner stable circular orbit (ISCO). However, in the neutron star accretor scenario, if the neutron star is weakly or non-magnetized, then the inner accretion flow will extend to the neutron star surface or the boundary layer and release energy in the form of additional harder emission component. In the highly magnetized neutron star case, the inner accretion flow truncates at the magnetospheric radius ($R_M$). There the emission from accretion column can produce harder emission component. However, the presence of significant outflowing wind can mask the additional harder spectral component from the line of sight. The inclination angle of the disk is also a contributing factor in shrouding this additional component. Hence, the two-thermal component is a representative of physical accretion flows in both black hole and neutron star system. We discuss these two scenarios in detail.

\subsection{Black hole model}
The bolometric luminosity of the hot \texttt{diskpbb} component is around $\rm L_{bol} = 4\pi D^2 F_{bol}  \simeq 1.5 \times 10^{40}$ \lumcgs. Such high luminosity is expected to be generated via super-Eddington accretion and thus further justifying that the temperature profile of the inner region of the disk diverges from standard thin disk and takes the form of $\rm T(r) \propto r^{-0.5}$. Here it is important to discuss that typically, for a spherical emitter, $\rm L_{bol} = 4\pi D^2 F_{bol}$ relation is justified. However, for accretion disks, it is shown that $\rm L_{bol} = (2\pi D^2/\cos\theta) ~F_{bol}$ is appropriate \citep{Fukue2000, Urquhart2016}, and the estimated luminosity will be dependent on the disk inclination angle. Nevertheless, it is common practice to estimate $\rm L_{bol}$ from $\rm 4\pi D^2 F_{bol}$, which is equivalent to the case of accretion disk at $60^{\circ}$ inclination. However, for super-critical disks, the self-occultation of the disk due to the geometrical thickness at a high inclination angle and self-irradiation further modifies this simple flux-luminosity relation \citep{Fukue2000}. However, for simplicity, we consider the  $\rm L_{bol} = (2\pi D^2/\cos\theta) ~F_{bol}$ relation in our work.

We can quantify some physical parameters from the spectral fitting. We calculate the inner radius of the disk from the best-fit normalization of the hot disk component ($\sim 4.45 \times 10^{-4}$). Here we assume a constant radius of the accretion disk. The inner disk radius is given by the form $\rm R_{in} = \xi \kappa^2 N^{0.5}D_{10}(\cos\theta)^{-0.5}$ km, where $\rm N$ is the normalization, $\rm \theta$ is the disk inclination, $\rm D_{10}$ is the distance to the source in $10$ kpc unit, $\rm \xi$ is the geometric and $\rm \kappa$ is the color correction factor \citep{Kubota1998, Soria2015}. The inner radius and the mass of the black hole are related by the form $\rm R_{in} = 6\alpha \frac{GM}{c^2}$, for a Keplerian orbit, where $\alpha$ is a function of spin parameter and can take the value of $1$ for a non-rotating black hole, or $\sim 0.21$ for an extremely rotating (prograde) Kerr black hole with spin parameter $a^* \sim 0.998$ \citep{Bardeen1972, Thorne1974}. For the inner hot \texttt{diskpbb} component, using the $\xi$ and $\kappa$ factors as $0.353$ and $3$ \citep{Vierdayanti, Soria2015}, respectively, we get an inner radius of $\sim 50 (\cos\theta)^{-0.5}$ km. This would correspond to a $\sim 6$ \ms blackhole for a non-rotating and face-on system. The disk inclination angle will influence the mass estimate. However, it is generally understood that ultrasoft ULX sources are found to be high-inclination (close to edge-on) systems. On the contrary, the ULXs where a hot inner disk component is visible are mostly low inclination systems \citep{Gu2016}. Thus, for a realistic disk inclination $<60^{\circ}$ for NGC 6946 X-1, the mass would be $< 10$ \ms for a non-rotating black hole. Simulations for super-critical accretion onto black holes have shown different perspective of beaming effect in generating the hard X-ray emission in ULXs. For example, \citealt{Jiang2014} found that the hard X-ray emissions from the central region is broadly isotropic, whereas, on the contrary, \citealt{Sadowski2016} found that there is a significant beaming along the polar axis.

\begin{figure*}[t]
\centering
    \includegraphics[width=0.45\textwidth, keepaspectratio]{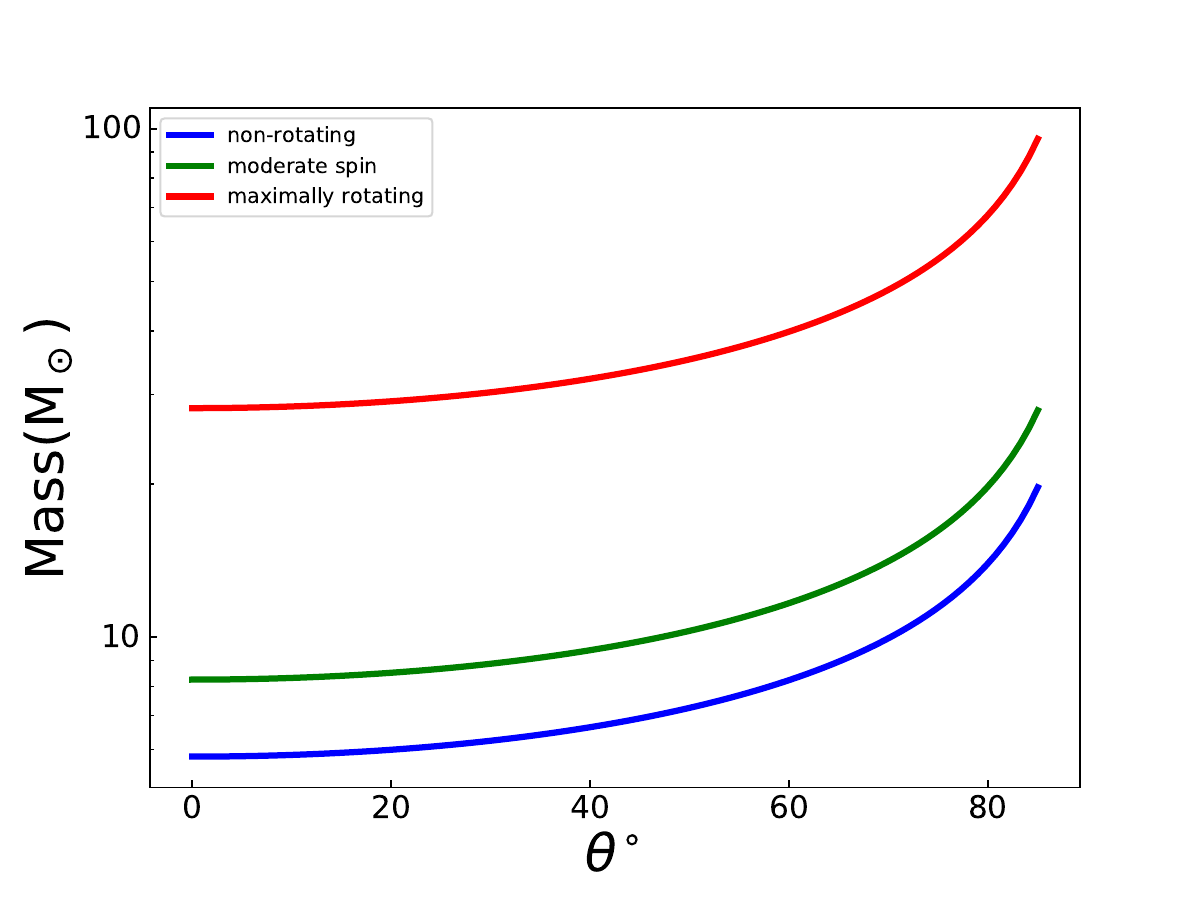}
    \includegraphics[width=0.45\textwidth, keepaspectratio]{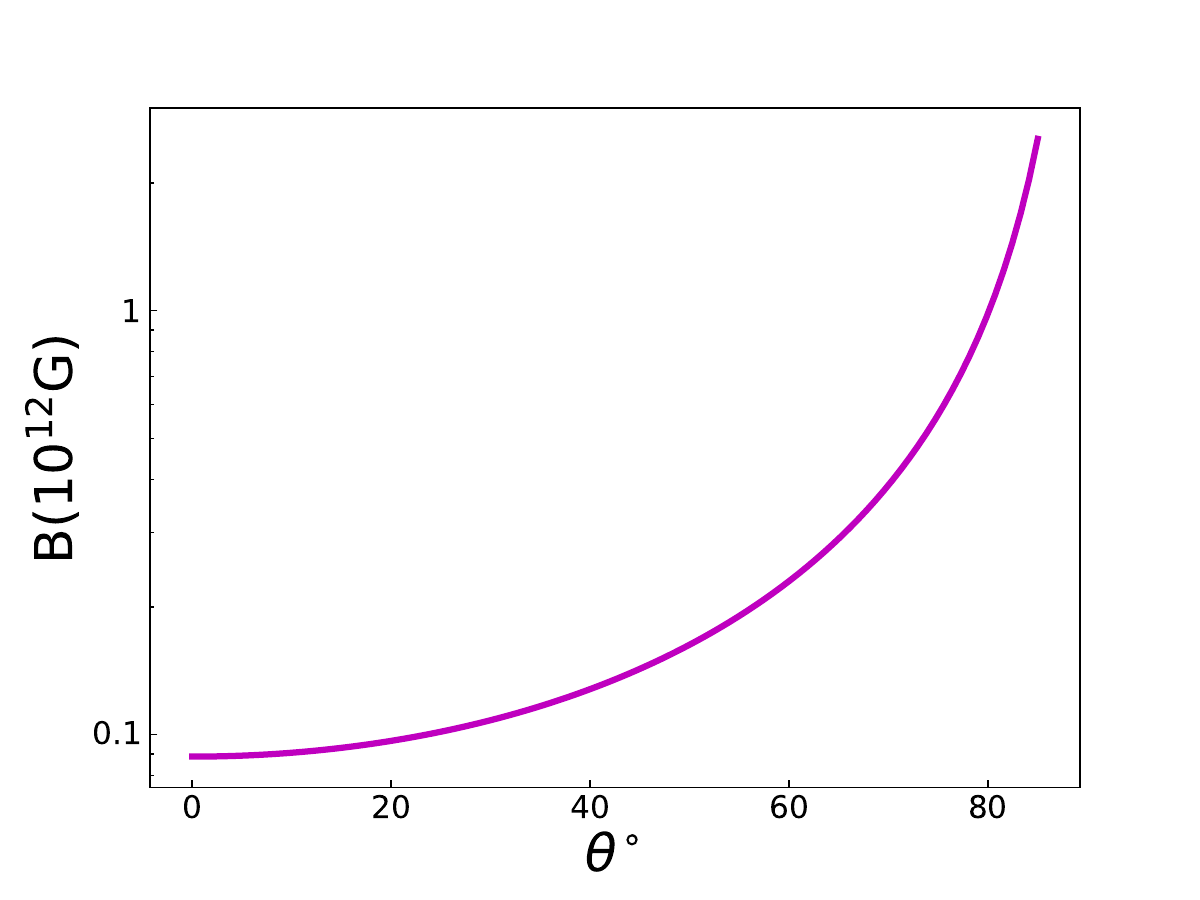}
\caption{Left: Variation of the estimated mass of the black hole as a function of disk inclination angle for three different spins of the black hole. The non-rotating black hole scenario of $a^*=0$, a moderate spin of $a^*=0.5$, and a maximally rotating case of $a^*=0.998$ are shown in the figure. Right: Dependency of estimated magnetic field strength for a neutron star system on the disk inclination angle.
\label{fig:mass_B_plot}}
\end{figure*}

In figure \ref{fig:mass_B_plot} (left), we show the dependency of mass estimate on the disk inclination with different black hole spin. Caution is necessary because this mass estimate depends on the assumption of a Keplerian orbit of constant radius, which might be different in reality depending on the geometry of the disk.

\subsection{Neutron star model} \label{sec:NS_discussion}
In a neutron star system, the inner hot disk can be interpreted to be truncated by a high magnetic field at the magnetospheric radius ($\rm R_M$). Typically, $\rm R_M = 7\times 10^7 \Lambda m^{1/7} R_6^{10/7} B_{12}^{4/7} L_{39}^{-2/7}$ cm, where $\rm m = M/\ms$ is the neutron star mass in solar mass units, $\rm B_{12}=B/10^{12}~ G$, $\rm L_{39} = L/10^{39}~ \lumcgs$, $\rm R_6 = R/10^6$ cm and for disk accretion $\rm \Lambda \sim 0.5$ \citep{Mushtukov2017}. This is assumed for a dipole structure magnetic field lines around the neutron star. If the hot inner region of the disk is coming from the \texttt{diskpbb} component, then we can estimate typical magnetic field strength equating the inner radius with the $\rm R_M$. We find that for a $\rm 1.4 ~\ms$ neutron star and radius of $10^6$ cm, the magnetic field $\rm B\sim 2 \times 10^{11}$ G for a $60^{\circ}$ inclined disk with bolometric inner disk luminosity $\sim 1.5 \times 10^{40}$ \lumcgs. We plot the estimated magnetic field strength and corresponding disk inclination angle in figure \ref{fig:mass_B_plot} (right). In the plot, the bolometric luminosity depends on the angle as $\sim 1.5 \times 10^{40} ~\lumcgs /(2\cos\theta)$. Since X-1 is not expected to be an extremely high inclination system, this simple angle dependence of luminosity is viable. In realistic inclination, $<60^{\circ}$ case, the strength of the field is $B \lesssim 2 \times 10^{11}$ G. Hence, if NGC 6946 X-1 is a neutron star system, then it probably hosts a moderately magnetized neutron star core.

Several studies have explored different spectral models to justify the neutron star scenario in ULX systems. One characteristic feature is to study the powerlaw model with a high-energy exponential cutoff. The spectral cutoff in neutron stars are often identified as the emission from accretion column \citep{Walton2018apr}. \citealt{Pintore2017}, investigated the pulsator-like spectra in ULXs by characterizing them with \texttt{highecut*powerlaw} model. NGC 6946 X-1 was studied in the sample with an archival \xmm\ only data. Along with an exponentially cutoff powerlaw continuum, a soft blackbody excess and the $\sim 1$ keV feature were detected. However, we find that the latest broadband data is sufficiently well fitted with a \texttt{gaussian+highecut*powerlaw} model in our analysis. Nevertheless, we study a crucial comparison with the pulsator-like spectral model and estimate the hardness and softness ratio as defined in \citealt{Pintore2017}. We estimate the total unabsorbed flux in $6.0-30.0$ keV, $4.0-6.0$ keV, and $2.0-4.0$ keV bands. We find that hardness $\frac{\rm F_x(6.0-30.0)}{F_x(4.0-6.0)} \sim 1$ and softness $\frac{\rm F_x(2.0-4.0)}{F_x(4.0-6.0)} \sim 2$. This result remains similar to the finding in \citealt{Pintore2017}. Typically neutron star systems are expected to have more hardness and lesser softness values. This study can indicate that if X-1 is a neutron star system, then it is not highly magnetized, consistent with our estimates, thus making the source less hard. Again, the inclination angle of the disk and the presence of optically thick wind also play a role in determining the hardness of the source.

\subsection{Accretion onto X-1}
The spectral properties of NGC 6946 X-1 provide evidence of super-critical accretion onto a low massive black hole or a moderately magnetized neutron star. Also, an apparent signature of the optically thick wind is expected in such a super-critical accretion scenario. This helps us constrain some physical accretion parameters from a realistic point of view.

For a super-critical accretion disk, the total accretion luminosity can be related to the Eddington factor ( $\rm \dot{m}_0 = \dot{M}_0/\dot{M}_{Edd}$) by  the following relation (see \citealt{Shakura1973}),

\begin{equation}
    \rm  L \simeq L_{Edd}[1+ln~\dot{m}_0] \label{lum_rate_relation}
\end{equation}

In our estimates, we make some assumptions. The luminosity we assume is the total bolometric unabsorbed luminosity. To estimate the luminosity, we have assumed a disk inclination angle of $60^{\circ}$, so that $\rm L_{bol} = 4\pi D^2 F_{bol} = 1.9 \times 10^{40} ~\lumcgs$, which is equivalent to the isotropic apparent luminosity relevant for Eq. \ref{lum_rate_relation}.

However, the important factor which constrains the $\rm \dot{m}_0$ comes from the beaming \citep{King2009, King2016}. An approximate beaming relation gives $\rm b \simeq 73/\dot{m}_0^2$ and the Eddington luminosity is given by $\rm L_{Edd} = 1.5 \times 10^{38} m_1 ~\lumcgs$ \citep{Poutanen2007}. Then the relation becomes,

\begin{equation}
    \rm \frac{m_1}{L_{40}} \simeq \frac{4900}{\dot{m}^2_0(1+ln~\dot{m}_0)},
\end{equation}

where the accretor mass $\rm m_1 = M/\ms$ and the luminosity determine the accretion rate.

The spherization radius $\rm R_{sph}$ for a super-Eddington disk is determined by the accretion rate by \citep{Shakura1973, Begelman, King2009},

\begin{equation}
    \rm R_{sph} \simeq \frac{27}{4} \dot{m}_0 \frac{2 G M}{c^2}
\end{equation}

For the bolometric total luminosity of $1.9 \times 10^{40}$ \lumcgs, if we assume a neutron star of $1.4$ \ms, then $\rm \dot{m}_0 \simeq 38$ or in the case of a $\sim 10$ \ms black hole, $\rm \dot{m}_0 \simeq 16$.

Eddington accretion rate ($\rm \dot{M}_{Edd} \simeq 2 \times 10^{18} m_1 ~g ~s^{-1} $) of a neutron star is $\sim 4.4 \times 10^{-8}$ \ms yr$^{-1}$. This gives the accretion rate of X-1 to be $\rm \dot{M}_0 = 1.7 \times 10^{-6}$ \ms yr $^{-1}$, if it is a neutron star. On the other hand, if it is a $10$ \ms black hole, then $\rm \dot{M}_{Edd} \sim 3.2 \times 10^{-7}$ \ms yr$^{-1}$ and $\rm \dot{M}_0$ for X-1 is $\sim 5 \times 10^{-6}$ \ms yr $^{-1}$.

The spherization radius $\rm R_{sph}$ is $\sim 1 \times 10^8$ cm for a neutron star system, and $\sim 3.2 \times 10^8$ cm for a $10$ \ms black hole. Now, as we have assumed the disk inclination of $60^{\circ}$, the inner radius from the hot disk component is $\rm R_{in} \sim 70 ~km \sim 7 \times 10^6$ cm. If we assume $\rm R_M \sim R_{in}$, then the magnetospheric radius is less than the $\rm R_{sph}$ of the neutron star estimate, i.e., $\rm R_M < R_{sph}$.

In the scenario of launching of optically thick wind down the $\rm R_{sph}$ for a neutron star, the condition $\rm R_M < R_{sph}$ is self-consistent \citep{King2016}. It is interesting to put a stringent constraint on the relation between $\rm \dot{m}_0$, the magnetic field strength $\rm B_{12}$, and disk inclination $\theta$ from the self-consistent condition. If we assume that $\rm R_M$ is estimated from the inner disk luminosity (which depends on the inclination angle of the disk), then a simple estimate of the relation would be,

\begin{equation}\label{eq:m_dot_gen_rel}
   \rm  \dot{m}_0 > 16 ~(L^{sph}_{39})^{-2/7} ~B_{12}^{4/7} ~(\cos\theta)^{2/7} ,
\end{equation}

where, $\rm L^{sph}_{39}$ is the apparent spherical luminosity from the disk in units of $10^{39} ~\lumcgs$. Thus, the relation for NGC 6946 X-1, would become,

\begin{equation}\label{eq:m_dot_6946_rel}
   \rm  \dot{m}_0 > 7 ~B_{12}^{4/7} ~(\cos\theta)^{2/7} ,
\end{equation}

which is satisfied by the estimated accretion rate, and magnetic field for relevant realistic disk inclination angles. It is important to note that Eq. \ref{eq:m_dot_gen_rel} is a general relation, whereas Eq. \ref{eq:m_dot_6946_rel} gives the specific condition for the source NGC 6946 X-1.

Non-detection of pulsation in this source has one implication that $\rm R_M$ is much smaller than the estimated $\rm R_{in}$, and thus the pulsation is diluted to be detected \citep{Walton2018apr}. In that case, the estimated magnetic field in \S~\ref{sec:NS_discussion} could be further weaker with lesser $\rm R_M$. Thus, a necessary condition for a neutron star ULX to be detected as a pulsar is $\rm R_M \simeq R_{sph}$ and as a consequence, these systems must possess high spin up rates as discussed in \citet{King2017}. In the case of a $10$ \ms black hole also, $\rm R_{in}$ which may be comparable to the inner stable circular orbit radius ($\rm R_{ISCO}$) is much smaller than the $\rm R_{sph}$.

To summarize, NGC 6946 X-1 is found to be a persistent soft ultraluminous X-ray source. Detection of spectral curvature, presence of wind/outflow and high bolometric luminosity of a hot slim accretion disk scenario prefers super-Eddington accretion onto a stellar mass compact object. If the host is a non-rotating black hole, the mass would be $<10$ \ms, or the ULX can host a moderately magnetized neutron star. The estimates of physical length scales are consistent with a geometry where the disk height is extended down the spherization radius and gets truncated at the inner radius as the inner stable circular orbit of a black hole or magnetospheric radius of a neutron star.

%% Also note that the akcnowlodgment environment does not support long amounts of text. If you have a lot of people and institutions to acknowledge, do not use this command. Instead, create a new \section{Acknowledgments}.
\begin{acknowledgments}
The authors would like to thank the referee for providing valuable suggestions to improve the manuscript further. The scientific results of this article have used archival data (available at the High Energy Astrophysics Science Archive Research Center (HEASARC)) obtained with \xmm, an ESA science mission with instruments and contributions directly funded by ESA member states and NASA. This research has also utilized archival data (HEASARC) obtained with \nustar, a project led by Caltech, funded by NASA, and managed by the NASA Jet Propulsion Laboratory (JPL), and has made use of the NuSTAR Data Analysis Software (\texttt{NuSTARDAS}) jointly developed by the ASI Space Science Data Centre (SSDC, Italy) and the California Institute of Technology (Caltech, USA).
\end{acknowledgments}


\vspace{5mm}
\facilities{\xmm; \citet{XMM2001}, \nustar ; \citealt{NuSTAR} }


\software{HEASOFT (\url{https://heasarc.gsfc.nasa.gov/docs/software/heasoft/}; \citet{Heasoft2014}), \xmm\ SAS (\url{https://www.cosmos.esa.int/web/xmm-newton/sas}; \citet{Gabriel2004}), HENDRICS (\url{https://hendrics.stingray.science/en/latest/}; \citet{HENDRICS}), STINGRAY (\url{https://docs.stingray.science}; \citet{Stingray})
          }



%\appendix


\bibliography{NGC6946X1_ApJ}{}
\bibliographystyle{aasjournal}



\end{document}